\documentclass[twocolumn]{aastex63}
\accepted{August 14, 2021}
\submitjournal{ApJ}

\shorttitle{Capturing the physics of MaNGA galaxies with self-supervised Machine Learning}
\shortauthors{Regina Sarmiento et al.}
\graphicspath{{./}{figures/}}

\usepackage{lineno}
\usepackage{graphicx}
\usepackage{amssymb}
\usepackage{bm}
\usepackage{comment}
\newcommand{\cmmnt}[1]{}
\usepackage[normalem]{ulem}
\usepackage{xcolor}
\usepackage{txfonts}
\begin{document}

   \title{Capturing the physics of MaNGA galaxies with self-supervised Machine Learning}

    \correspondingauthor{Regina Sarmiento}
    \email{regina.sarmiento@iac.es}

   \author[0000-0002-3803-6952]{Regina Sarmiento}
   \affiliation{Instituto de Astrof{\'i}sica de Canarias (IAC)
   La Laguna, 38205, Spain}
   \affiliation{Departamento de Astrof{\'i}sica, Universidad de La Laguna (ULL) E-38200, La Laguna, Spain}
   
   \author{Marc Huertas-Company}
   \affiliation{Instituto de Astrof{\'i}sica de Canarias (IAC)
   La Laguna, 38205, Spain}
   \affiliation{Departamento de Astrof{\'i}sica, Universidad de La Laguna (ULL) E-38200, La Laguna, Spain}
   \affiliation{LERMA, Observatoire de Paris, PSL Research University, CNRS, Sorbonne Universit{\'e}s, UPMC Univ. Paris 06,F-75014 Paris, France}
   
   \author[0000-0003-1643-0024]{Johan H. Knapen}
   \affiliation{Instituto de Astrof{\'i}sica de Canarias (IAC)
   La Laguna, 38205, Spain}
   \affiliation{Departamento de Astrof{\'i}sica, Universidad de La Laguna (ULL) E-38200, La Laguna, Spain}  
   
   \author{Sebasti{\'a}n F. S{\'a}nchez}
   \affiliation{Instituto de Astronom{\'i}a, Universidad Nacional Aut{\'o}noma de Mexico}
   
   \author{Helena Dom{\'i}nguez S{\'a}nchez}
   \affiliation{Department of Physics and Astronomy, University of Pennsylvania, Philadelphia, PA 19104, USA}
   \affiliation{Institute of Space Sciences (ICE, CSIC), Campus UAB, Carrer de Magrans, E-08193 Barcelona, Spain}   
   
   \author{Niv Drory}
   \affiliation{McDonald Observatory, The University of Texas at Austin, 1 University Station, Austin, TX 78712, USA}   
   
   \author{Jesus Falc{\'o}n-Barroso}
   \affiliation{Instituto de Astrof{\'i}sica de Canarias (IAC)
   La Laguna, 38205, Spain}
   \affiliation{Departamento de Astrof{\'i}sica, Universidad de La Laguna (ULL) E-38200, La Laguna, Spain}





  \begin{abstract}
   
   As available data sets grow in size and complexity, advanced visualization tools enabling their exploration and analysis become more important. In modern astronomy, integral field spectroscopic galaxy surveys are a clear example of increasing dimensionality and complexity of datasets, which challenge the traditional methods used to extract the physical information they contain. We present the use of a novel self-supervised Machine Learning method to visualize the multi-dimensional information on stellar population and kinematics in the MaNGA survey in a two dimensional plane. Our framework is insensitive to non-physical properties such as the size of integral field unit (IFU) and is therefore able to order galaxies according to their resolved physical properties. Using the extracted representations, we study how galaxies distribute based on their resolved and global physical properties. We show that even when using exclusively information about the internal structure, galaxies naturally cluster into two well-known categories from a purely data driven perspective: rotating main-sequence disks and massive slow rotators, hence confirming distinct assembly channels. Low-mass rotation-dominated quenched galaxies appear only as a third cluster if information about the integrated physical properties is preserved, suggesting a mixture of assembly processes for these galaxies without any particular signature in their internal kinematics that  distinguishes them from the two main groups. The framework for data exploration is publicly released with this publication, ready to be used with the MaNGA or other integral field data sets.

\end{abstract}


%

\section{Introduction}\label{sec:intro}

Over the past decades, the sizes of available data sets in astronomy have grown by orders of magnitude (e.g. Sloan Digital Sky Survey (SDSS; \citealt{SDSS-Gunn}), COSMOS (\citealt{cosmos})) and the trend is expected to accelerate in this new decade (e.g. Vera C. Rubin Observatory Legacy Survey of Space and Time (LSST), \citealt{LSST-overview}; {\it Euclid}, \citealt{euclid}; {\it Nancy Grace Roman Space Telescope}, \citealt{Roman-overview}). As data sets grow not only in size but also in complexity, efficient exploratory methods are needed to boost discovery. Dimensionality reduction techniques are vital in this respect (\citealt{Dalya-sequencer}, \citealt{Portillo-SDSS-VAE}). However, astronomy data present some specific problems such as noise and complex selection effects which challenge the use of standard techniques out of the box. This is a field where recent advances in Machine Learning can undoubtedly be of major utility.

We explore in this work the application of self-supervised Contrastive Learning as a tool to explore multi-dimensional astronomy data. We employ it to analyse maps derived from the SDSS-IV (\citealt{MaNGA-Blanton-2017}) Mapping Nearby Galaxies at APO (MaNGA; \citealt{MaNGA-Bundy-2015}) survey of close to $10^4$ nearby galaxies with dedicated integral field spectroscopic fiber bundles. This is a particularly complex survey in which multiple sampling sizes are used and it presents a complex selection function. 

Contrastive Learning has rapidly emerged in the Machine Learning community as a powerful solution to find meaningful representations of data. As opposed to more classical dimensionality reduction approaches such as principal component analysis (PCA; \citealt{PCA-1901}) or Auto Encoders (\citealt{autoencoders}) which aim at projecting data into a low dimensional space to minimize the reconstruction error, contrastive learning maximizes the agreement between different transformations of the same data point with no reconstruction involved in the process.  This is particularly well adapted to datasets where observational effects can make objects with similar physical properties appear very different (e.g. due to PSF variations, sampling, S/N etc). The representations learned can not only be used for exploration but also for downstream tasks such as anomaly detection or classification  with the advantage of reducing the sizes of the training sets. A major breakthrough of last year was indeed that a supervised classification based on contrastive learning representations achieved better accuracy than a purely convolutional neural network-based supervised classification on the ImageNet database (\citealt{chen2020big}). Even more recently, \cite{selfsup-galaxy} proved a similar behavior in astronomical data. They analyzed multi-band optical images from the SDSS and found that photometric redshift and galaxy morphology classifications can be estimated with the same accuracy as a completely supervised approach but using half the number of labels during training.

In this work, we use contrastive learning to explore the multi-dimensional kinematic and stellar population maps of galaxies derived from the MaNGA dataset. This is the first time contrastive learning has been applied to integral field spectroscopy data. We use prior knowledge on observational biases in order to remove these artificial features and extract physical information from the data. We aim to explore to what extent physically relevant parameters and relations can be reproduced or uncovered automatically. Testing such algorithms on MaNGA data allows us to compare our findings with results obtained within the survey project using more traditional analysis tools. This is an important validation step which must be taken before we can consider using self-supervised analysis on larger, more complicated, or newer data sets.

This paper is organised as follows. In Section~2 we describe the data we used, and in Section~3 the machine learning software applied. Section~4 describes our results in terms of visualisation, regression, feature decomposition and clustering. Implications of our work are discussed in Section~5, and summarised in Section~6.

\section{Data}\label{sec:data}

We use data products from the MaNGA survey (\citealt{MaNGA-Bundy-2015}, \citealt{Drory-2015-MangaMandatory}, \citealt{Aguado-DR15}). MaNGA is one of the three core surveys in the fourth-generation SDSS (SDSS-IV; \citealt{MaNGA-Blanton-2017}, \citealt{sdss-boss-mandatory}). It uses 17 fiber-bundle integral field units (IFUs) that vary in diameter from 12\,arcsec (19 fibers) to 32\,arcsec (127 fibers of 2\,arcsec diameter) to observe a total of $10^4$ nearby galaxies (\citealt{Law-2015-MangaMandatory}, \citealt{Yan-2016-MangaMandatory}, \citealt{Wake_2017}), constituting a representative sample of all types of nearby galaxies in a redshift range $0.01<\,z\,<\,0.15$. Data-cubes are obtained with IFUs of 19, 37, 61, 91 and 127 fibers, matched to the galaxy size, with spectra covering the range 360$-$1000nm in wavelength at a spectral resolution  $R\sim$2000) (\citealt{Law-2016-MangaMandatory}, \citealt{Yan-2016-MangaMandatoryB}).

The data set used in this work consists of post-processed maps derived from MaNGA MPL-10 data release using the Pipe3D pipeline (v3.0.1b) (\citealt{sanchez2015pipe3d}, \citealt{sanchez2016pipe3d}). Pipe3D performs a spectral fit of a combination of three synthetic components: stellar populations, dust and gas. The pipeline produces maps by deriving the properties of averaged stellar population. To test the performance of the algorithm, we use integrated magnitudes derived from the post-processed MaNGA data cubes, which are included in the DR14Pipe3D primary catalog.

The input data for which we aim at finding low-dimensional representations consists of five binned ($2\times2$) and stacked hexagonal maps for each galaxy, namely the $V-$band reconstructed images, luminosity-weighted age and metallicity maps, and radial velocity and velocity dispersion maps. Previous to the binnig, the pixels with a signal to noise ratio lower than $3$ in the median intensity maps are set to zero. As the MaNGA IFUs have different numbers of fibers, the sizes of the resulting maps vary as well. Since convolutional neural networks (CNNs) typically require input maps of a fixed shape, we zero-padded the hexagonal maps to the same shape and size $32\times32$. In order to deal with the different dynamic ranges, we scaled all maps linearly to the range [0,1] except for the V-band image, to which we applied logarithmic and square root functions previous to the linear normalization. Two different approaches will be analyzed regarding the normalization. On the one hand, we consider a normalization that is equivalent to a change of units that is uniformly applied to all the galaxies (we will refer to it as \emph{relative norm}), as we aim to analyze the relative physical magnitudes among the galaxies. In this case, all maps have values between $0-1$, but not necessarily span the complete range. On the other hand, a normalization that forces each galaxy map to span the [0,1] range is used to evaluate the effect of the spatial structures of the galaxies (the \emph{individual norm}).

Our final dataset consists of 9507 MaNGA galaxies, of which 1099, 2235, 2235, 1122, and 2816 were observed with a 19, 37, 61, 91, and 127-fiber bundle, respectively.

\section{Deriving meaningful representations of MaNGA galaxies}\label{sec:reps}

\subsection{Contrastive model}

We use the Simple framework for Contrastive Learning of visual Representations (SimCLR; \citealt{chen2020simple}). SimCLR is a contrastive learning framework to extract meaningful representations of the data. More precisely, we use the model in Tensorflow \citep{tensorflow2015-whitepaper} and follow the findings on optimization by \cite{chen2020simple} with minor changes in the architecture. 

This algorithm optimises representations that keep the semantic, or physical, information of the data by maximizing the agreement between the representations of a data object and its transformed pair $(\boldsymbol{x}, \boldsymbol{x}^{\rm T})$. This is particularly beneficial when similar objects appear to be different due to observational biases. By selecting transformations that simulate these observational biases the model is encouraged to neglect these differences while capturing the common remaining features, which correspond solely to the galaxy. Likewise, agreement is minimized when representations originate in different galaxy maps. This way, the program is self-taught to recognize meaningful patterns present in the data. 

More specifically, the model consists of a CNN block followed by a set of fully connected layers after which the contrastive loss $l$ is computed. The CNN block acts as a base encoder that outputs the representations, while the fully connected layers project the representations to the comparison space. We will refer to these last fully connected layers as head projection function. While both components are trained together, the head projection is removed to obtain the final galaxy representations. \cite{chen2020simple} found that this configuration yields better representations in the previous layers.

The CNN base encoder consists of four convolutional layers with kernel sizes 5, 3, 3, 3 and 128, 256, 512, 1024 filters each. Max-pooling layers and Exponential Linear Unit activation functions \citep{ELU-clevert2016} are used between the convolutional layers. The output of this set of layers encodes the array-like representations of the galaxies of 1024 features each. 

The projection head is composed of three fully connected layers (512, 128 and 64 neurons each). This non-linear function takes the galaxy representations to a latent space where the contrastive loss is computed.

The contrastive loss is calculated for each $L^2$-normalized representation $\boldsymbol{z}_i$ in the contrastive space with the following equation: 
\begin{equation}
    l_{i,j} = -\log \frac{\exp(\langle \boldsymbol{z}_i, \boldsymbol{z}_j\rangle/h)} {\Sigma_{k=1, k \neq i}^{2N}\,\,\exp(\langle \boldsymbol{z}_i, \boldsymbol{z}_k\rangle/h)},
    \label{contrastive_loss}
\end{equation}

where $\langle,\rangle$ is the dot product, $h$ is a temperature factor that regulates the distribution of the output representations by decreasing or increasing its concentration (\citealt{wu2018}, \citealt{hinton2015}), in our case, $h$ is set to $0.5$. $\boldsymbol{z}_j$ denotes the representation of the augmented view of $\boldsymbol{x}_i$, while $k$ sweeps the other $2N-1$ representations in the batch.

This loss, first introduced in \cite{loss-paper}, does not assume a prior distribution of the representation space and therefore can be used to learn representations of both continuous or discrete data sets. The model benefits from large number of negative examples and long training times \citep{chen2020simple}. Our model was trained using a batch size of 1024 during 500 epochs, which represents $40$ minutes of computing time on one GeForce RTX 2080 Ti GPU with NVIDIA V-\,$450.102.04$.

\subsection{Augmentations}

Data augmentation is a common practice in Machine Learning that consists of applying transformations to the input data during the training step. The goal is to create different views of the objects while keeping their semantic information. In a supervised approach, data augmentation can be used to enlarge the training sample, since the different views of the data will have the same labels. In the contrastive learning framework, augmentations are essential to create positive pairs of objects, these are the transformed images that originate from the same image. 

\begin{figure}
    \centering
    \includegraphics[width=\hsize]{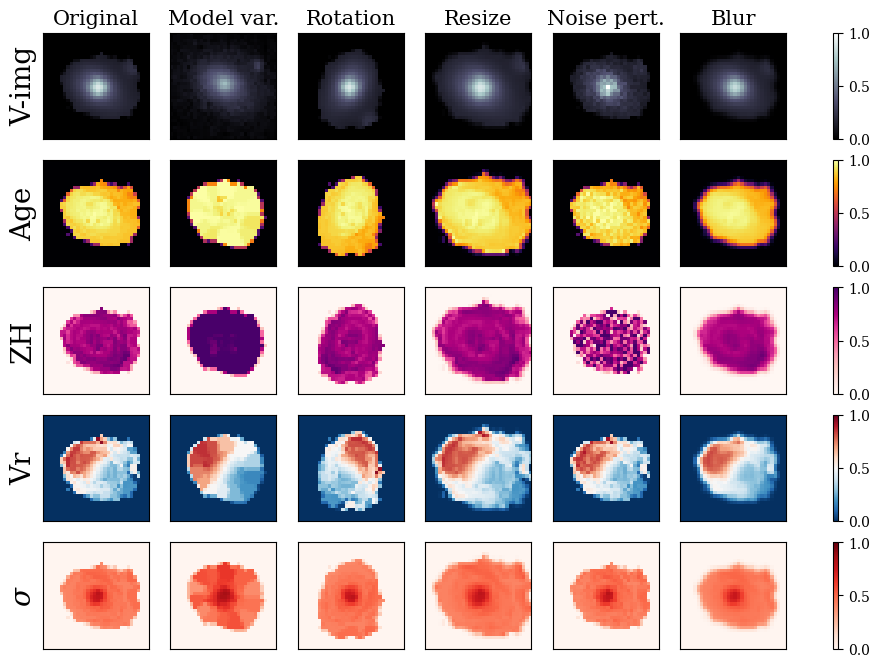}
    \caption{Augmentations used during training and their effect on the different input channels of an example galaxy. The leftmost column shows the original five maps of the example galaxy in adimensional units, from top to bottom: $V-$band reconstructed image, luminosity weighted age and metallicity, radial velocity and velocity dispersion. The columns to the right show the resulting maps after each transformation has been applied. Namely, model variation, rotation, resize, noise perturbation and blur.}
    \label{augmentations}
\end{figure}

When generating meaningful representations of the data in our machine learning approach, we seek that galaxies that are similar in intrinsic physical parameters will appear close to each other in the representation space.  A standard but non-optimal feature decomposition of the data may reproduce features that are intrinsic to the data, but not representative of the galaxies. The possible origin of such features is diverse, and may be related to the survey design, to instrumental effects, to the pipeline used to process the data, or top observational constraints such as orientation or apparent size of the target. Therefore, the augmentation functions we used during the training of the CNN were designed for the MaNGA data set and with the purpose of mitigating non-physical dependencies as detailed below and illustrated in Figure~\ref{augmentations}.

\begin{itemize}

\item Model variation: To prevent the model from learning modelling systematics that could be introduced by the pipeline, a transformation that replaces the input set of maps by a set of differently modelled maps is included. With this transformation, the kinematic maps are replaced by those obtained from the MaNGA Data Analysis Pipeline (DAP) which uses a Voronoi binning to ensure a S/N=10 in the g-band \citep{voronoi-bin} and MILES template library for the stellar continuum fitting (\citealt{miles1}, \citealt{miles2}). The luminosity weighted age and metallicity maps are replaced by the FIREFLY catalog (\citealt{firefly1}, \citealt{firefly2}) analogs. The V-band reconstructed image is replaced by the scale matched SDSS r-band photometric image\footnote{SDSS r-band photometric image as in: \url{http://skyserver.sdss.org/ImgCutoutDR7/ImgCutout.asmx}}, which represents a change of view of the galaxy morphological features rather than a modelling variation.

\item Random flip and rotation (0$^{\circ}$, 90$^{\circ}$, 180$^{\circ}$, 270$^{\circ}$): Since there is no preference in the orientation of the galaxies, the model should be invariant to the galaxy position. We quantify this dependence with the galaxies position angle and rotation angle. While the first is used as listed in the DR14Pipe3D catalog, the second was estimated from each velocity map as the perpendicular angle to the direction defined by the mean position of the negative pixels and the mean position of the positive pixels.

\item Noise perturbation: The data is affected by errors inherent to the instrument and the post-processing of the observations. This transformation aims to compensate for measurement uncertainties by shifting the values of the map pixels. The transformation consists of multiplying each value by a random normal factor $\mathcal{N}(1, \sigma$), with $\sigma$ adjusted to the expected error for each pixel of each map. $\sigma$ maps are calculated as the averaged signal-to-noise maps for each channel\footnote{As error maps for the V-band reconstructed image were not available, a $\sigma=0.1$ has been assumed for this channel.}.

\item Resize: Since the model uses a fixed size for the input, IFUs with a smaller number of fibers will be more intensively zero-padded than IFUs with more fibers. Even though the IFU size used for each galaxy is linked to its apparent size, the MaNGA survey has an uneven sampling of galaxy angular sizes to IFU size. The IFU size is chosen for each galaxy by optimizing a $1.5 R_{\rm e}$ angular coverage of the target object in the Primary sample, and $2.5 R_{\rm e}$ in the Secondary sample \citep{MaNGA-Bundy-2015}, which means that the ratio $D_{\rm IFU}[{\rm arcsec}]/R_{\rm e}[{\rm arcsec}]$ ranges from $1.5$ to $2.5$ when combining Primary and Secondary samples. To prevent the algorithm from considering the IFU size and resolution as relevant features, the resize transformation randomly enlarges or reduces the image size to up to 25\% of the side size of the original frame. Cubic interpolation is used.

\item Gaussian blur: A 2D Gaussian kernel of $3\times 3$ pixels\,$^2$ and $\sigma$\,[pixels] from $\mathcal{U}[0.1, 2)$ is convolved with each layer of the data input. This encourages the model to become invariant to different resolutions and sharp artificial features in the data maps. 

\end{itemize}

While Noise perturbation, Resize and Gaussian blur transformations have a $50\,\%$ probability of being applied to the raw data input copy, Random flip and rotation is always applied for a better performance. The Model variation augmentation is applied with a lower probability since the Voronoi tessellation of the stellar maps could affect the performance of the kernels of the CNN. Therefore, this augmentation randomly replaces $25\,\%$ of the copies.


\section{Removing instrumental biases: comparison with PCA}\label{sec:res1}

We now assess how the learned representations correlate with known physical properties of the data. For this purpose, the input data normalized to keep relative values between galaxies is used. We first visually explore how galaxies are distributed in the representation space. We then illustrate a clustering application to characterize groups in specific regions of the representation space.

\subsection{Two-dimensional visualization of representations}\label{sec:2dvisual}

\begin{figure*}
   \centering
   \includegraphics[width=0.7\hsize]{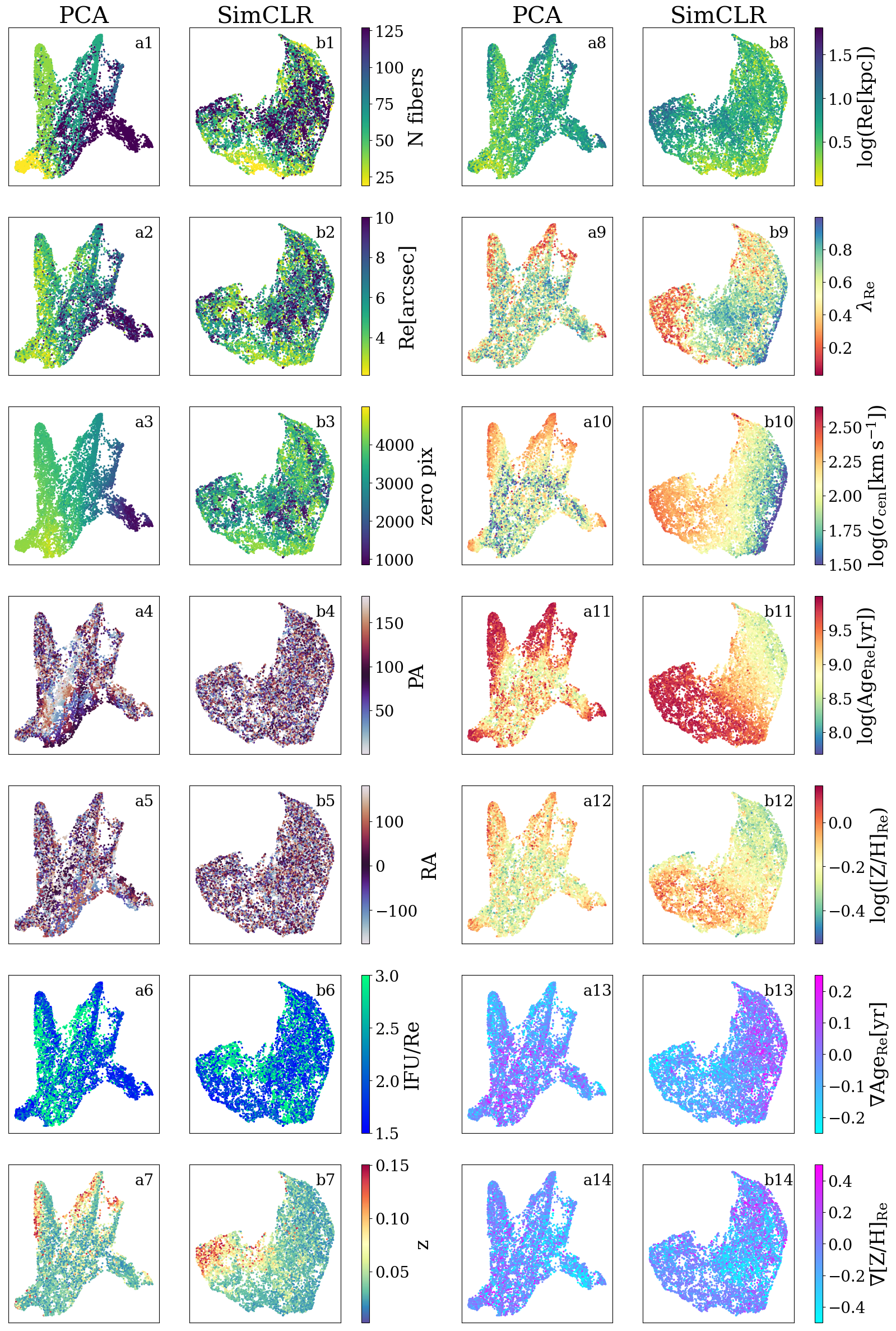}
      \caption{2D UMAP visualization of PCA decomposition (columns 1 and 3) and SimCLR representations (columns 2 and 4) color-coded according to different labels. The two left columns show dependencies with non-physical labels, from top to bottom: number of fibers in the IFU, galaxy angular size, number of zero pixels, position and rotation angle of the galaxy, angular coverage and redshift. While the two columns on the right are color-coded according to integrated parameters derived from the input maps, from top to bottom: effective radius, specific angular momentum, central velocity dispersion, luminosity weighted age and metallicity and their gradients. Principal components are dominantly distributed in the 2D projection according to the number of fibers in their IFUs. The SimCLR representations show poorer correlations with undesired dependencies, while presenting smooth transitions with physical parameters.}
         \label{fig:2d_comparison}
\end{figure*}

In order to easily visualize the representation space we perform a dimensionality reduction of the representations of dimension 1024 to a 2D space using Uniform Manifold Approximation and Projection (UMAP; \citealt{umap}) and color-code using different physical and instrumental properties. This is shown in Fig.~\ref{fig:2d_comparison}. We emphasize that the UMAP representation is only used here for visualization purposes. The actual dimensionality of the representation space is higher than these two dimensions. Our goal is to understand how galaxies are organized in the representation space and more precisely whether non-physical dependencies have been properly removed. We therefore include, on the one hand, the parameters related to instrumental effects (leftmost and left-center panels in Fig.~\ref{fig:2d_comparison}), which are (top to bottom): number of fibers in the IFU, galaxy angular size ($R_{\rm e}{\rm [arcsec]}$), number of zero pixels, position and rotation angle of the galaxy (PA and RA, respectively), angular coverage ($D_{\rm IFU}[{\rm arcsec}]/R_{\rm e}[{\rm arcsec}]$) and redshift $z$ to account for physical resolution. On the other hand, we include a set of integrated physical properties derived from the input maps (right-center and rightmost panels), namely (top to bottom) effective radius ($R_{\rm e}$\,[kpc]), specific angular momentum within $1.5 R_{\rm e}$ ($\lambda_{R_{\rm e}}$), radial velocity dispersion at center of frame ($\sigma_{\rm cen}$), luminosity weighted age at $R_{\rm e}$ and luminosity weighted metallicity at $R_{\rm e}$ normalized by the solar metallicity ($[Z/H]_{R_{\rm e}}$). We also include the slopes of the gradients within $0.5$ to $2 R_{\rm e}$ of the last two parameters ($\nabla {\rm age}_{R_{\rm e}}$ and $\nabla [Z/H]_{R_{\rm e}}$).

To better quantify the efficiency of the removal of instrumental dependencies, we contrast representations obtained by a PCA with those obtained with the SimCLR algorithm. For the first set, the five MaNGA maps are decomposed to 1024 principal components, to match the SimCLR representations dimension. In all panels, the distribution of the 2D projection of the PCA feature decomposition, on the left (labelled with "a" in the top right corner), with the representations learned by contrastive learning, on the right (labelled with "b" in the top right corner). These projections are simply obtained by applying the UMAP to the features obtained with PCA, on the one hand, and with SimCLR, on the other. 

The principal components directly projected via UMAP are primarily organized based on instrumental and observational parameters (Fig. ~\ref{fig:2d_comparison}). The main features that drive the organization of the PCA space are indeed the number of zero pixels, the number of fibers on the IFU and the angular size of the galaxies (panels a1, a2 and a3). Although some slight dependencies on these parameters are recognizable in the SimCLR representation space, they are remarkably reduced (panels b1, b2 and b3). More precisely, we find that the principal components are locally organized according to the orientation parameters (position angle, PA, and rotation angle, RA) in panels a4 and a5, while the representations learned through the contrastive algorithm remove these orientation dependencies efficiently (panels b4 and b5). We notice, however, that the angular coverage dependencies were not completely removed (panel b6): an over-density of the MaNGA secondary sample is visible in the representation space. Panel b7 also shows a dependence of the representations with redshift, this could be due to the fact that the MaNGA survey has an overpopulation of massive galaxies toward higher $z$ \citep{Wake_2017}. When not considering a volume-limited sample, using $z$ to quantify non-physical dependencies could lead to the conclusion that representations depend on spatial resolution, whereas the features learnt actually reflect physical properties. Therefore, even if redshift indicates a physical resolution dependency, it is not a reliable indicator for this particular data set.

The rightmost column of Figure~\ref{fig:2d_comparison} shows that the SimCLR representation space smoothly transitions as a function of physical properties: age, metallicity and central velocity dispersion (panels b11, b12, b10). Furthermore, parameters that are less trivial to obtain from the maps, such as $\lambda_{R_{\rm e}}$ (b9) and $\nabla {\rm age}_{R_{\rm e}}$ (b13), are also learned by the model and show smooth transitions in the SimCLR representations. In general, the distribution of the principal components also shows a clear dependence on these parameters, but in this space galaxies that share the same properties tend to appear in separate areas of the diagram due to instrumental biases.

These results confirm that the SimCLR representations efficiently remove the complex instrumental and observational biases present in the MaNGA survey data and primarily organize galaxies based on physical parameters.

\subsection{Unsupervised Clustering of kinematics and stellar populations of galaxies}\label{sec:clustering}

Based on the encouraging results of the previous subsection, we further investigate how the sample of galaxies is clustered in the representation space by dividing the sample into groups using the $K$-means clustering algorithm (\citealt{scikit-learn}, \citealt{k-means-macqueen1967}) on the $1024$ dimensions of the representation space. $K$-means is a standard iterative algorithm to cluster unlabelled data by minimizing the sum of squared euclidean distances of the data-points within each cluster to their average position, or centroid. The number of clusters provided by the user will determine the number of centroids that will be initialized. We assume the same distance metric in the representation space as the one used to calculate the contrastive loss in the head projection space and therefore, normalize the representations with $L^2$ norm such that the euclidean and cosine (as $[1- \langle \boldsymbol{a}, \boldsymbol{b}\rangle]$, where $\boldsymbol{a}, \boldsymbol{b}$ are two array-like representations) distances between representations are equivalent. 

The resulting clusters represent a first-order division of representation space and will allow a further understanding of the underlying astrophysical properties that drive the space distribution by studying the galaxy groups. Since our representation space is continuous in nature, this exercise will not lead to a sharply defined, discrete, classification. Nevertheless, we define a criterion to obtain a repeatable and physically driven classification, allowing a further study of the galaxy groups thus identified. To define the number of clusters we require the division to be robust, i.e. it should be independent from the training run and initialization parameters. Additionally, the clusters must be independent of features that are not physically relevant.

To ensure a robust division, we train the SimCLR framework ten independent times such that the input maps are perturbed differently in each run. By construction, the features obtained in each run are not exactly the same since the training has some inherent stochastic behavior. Nevertheless, our assumption is that the representation spaces must encode the same information irrespectively of the initialization, such that neighboring galaxies in one run should remain neighbors in the other resulting spaces. We set the first run as the reference run and maximize matching clusters of the remaining runs with the linear assignment algorithm implemented in SciPy (\citealt{2020SciPy}, \citealt{lin-assig}). As the clusters of each run will not necessarily be identified under the same labels by the clustering algorithm, the linear assignment algorithm provides a new set of cluster labels such that the number of galaxies that fall in the same groups as the reference run is maximized. An agreement score is defined as the fraction of objects that are consistently placed in the same group in at least 90\% of the runs. Although this score might appear too restrictive when a large number of clusters is considered, it shows how reproducible the resulting groups are. We consider divisions to have good agreement when the agreement score is over $85$\,\%.

Finally, to account for undesired dependencies, e.g. on instrumental parameters, we compute the linear contribution of the non-physical parameters to the classifications found. We perform a linear regression fit of the parameters to a binary classification for each cluster. One class contains galaxies that belong to the given cluster while the other class is a random selection of non-member galaxies. The fit is presented with the same number of examples of both classes while a spare $20\%$ of the sample is used to obtain the validation accuracy. The resulting coefficient associated to each parameter is then normalized to indicate its contribution percentage to the classification. We reject clusters when the accuracy of this fit is higher than 65\%, in which case we assume that the cluster is the result of instrumental and/or non-physical effects.

For comparison purposes, we also perform a $K$-means clustering on a representation space derived through PCA on the five MaNGA maps. We note that the metrics chosen are compatible with the SimCLR framework, while this might not be the case for PCA. Therefore, a more favorable additional test is performed for the PCA decomposition. In this case, only the first 10 PCA parameters are selected and each parameter is scaled to have a $0$ mean and standard deviation of $1$.

The agreement score and the most accurate fit to the parameters that are not physically relevant are shown in Table~\ref{tab:clusters} for 14 divisions of the SimCLR representations and four divisions of the $1024$ and $10$ most relevant PCA features. For comparison, the relevance of physical and non-physical parameters in each cluster---computed through linear regression---are  also listed in the last two columns.

\begin{deluxetable}{lccc}[t!]
\tablenum{1}
\tablecaption{\label{tab:clusters}Cluster divisions performance.}
\tablewidth{0pt}
\tablehead{
\multicolumn4l{SimCLR} \\
\colhead{N clusters} & \colhead{\% Agree} & \colhead{Acc$_{\rm max}$Non-phys} & \colhead{Acc$_{\rm mean}$Phys}}
\startdata
\textbf{2} & \textbf{89.419 $\pm$ 0.006} & \textbf{64 $\pm$ 2} & 93.1 $\pm$ 0.5 \\
\textbf{3} & \textbf{93.63 $\pm$ 0.03} & \textbf{60.5 $\pm$ 0.5} & 89.3 $\pm$ 0.5 \\
4 & 62.5 $\pm$ 0.1 & \textbf{63 $\pm$ 2} & 86.4 $\pm$ 0.3 \\
5 & 84.43 $\pm$ 0.08 & 68 $\pm$ 1 & 85.5 $\pm$ 0.6 \\
6 & \textbf{85.04 $\pm$ 0.09} & 68 $\pm$ 1 & 84.1 $\pm$ 0.5 \\
7 & 60 $\pm$ 1 & 70 $\pm$ 1 & 83.1 $\pm$ 0.5 \\
8 & 58 $\pm$ 5 & 72 $\pm$ 1 & 82.4 $\pm$ 0.3 \\
9 & 56 $\pm$ 2 & 72.9$\pm$ 0.7 & 82.6 $\pm$ 0.5 \\
10 & 58 $\pm$ 2 & 71 $\pm$ 1 & 81.9 $\pm$ 0.5 \\
11 & 65 $\pm$ 3 & 71 $\pm$ 2 & 81.5 $\pm$ 0.9 \\
12 & 47 $\pm$ 2 & 71 $\pm$ 2 & 80.9 $\pm$ 0.7 \\
13 & 50 $\pm$ 3 & 72 $\pm$ 3 & 80.1 $\pm$ 0.7 \\
14 & 51 $\pm$ 5 & 73 $\pm$ 3 & 79.1 $\pm$ 0.8 \\
15 & 64 $\pm$ 8 & 75 $\pm$ 4 & 78.7 $\pm$ 0.9 \\
        \hline
        PCA 1024&&&\\
        N clusters&&Acc$_{\rm max}$Non-phys&Acc$_{\rm mean}$Phys\\
        \hline
        2 && 92.13 $\pm$ 0.09 & 72.28 $\pm$ 0.07 \\
        3 && 94.9 $\pm$ 0.8 & 71.3 $\pm$ 0.2 \\
        4 && 92.7 $\pm$ 0.7 & 64.7 $\pm$ 0.2 \\
        5 && 91.9 $\pm$ 0.3 & 69.5 $\pm$ 0.5 \\
        \hline
        PCA 10&&&\\
        N clusters&&Acc$_{\rm max}$Non-phys&Acc$_{\rm mean}$Phys\\
        \hline
        2 && 95 $\pm$ 3 & 72 $\pm$ 2 \\
        3 && 93 $\pm$ 2 & 67.4 $\pm$ 0.6 \\
        4 && 92 $\pm$ 1 & 68 $\pm$ 1 \\
        5 && 94.1 $\pm$ 0.5 & 64.2 $\pm$ 0.5\\
\enddata
\tablecomments{For each division into $N$ clusters an agreement score is calculated from ten different runs of the contrastive algorithm. Third column indicates the most accurate fit to non-physical parameters among the $N$ clusters, while the last column shows the average accuracy of the fit to integrated astrophysical parameters. Uncertainty for agreement score is estimated as the standard deviation due to K-means initialization with ten different seeds, while accuracy uncertainties were estimated as the standard deviation corresponding to five SimCLR independent runs with five different seeds each in the clustering step.}
\end{deluxetable}

In all cases, the SimCLR representations yield group divisions that are more dependent on physical features than on those that are not physically relevant, thus confirming the qualitative results of the previous section. In contrast, the divisions derived from the PCA are highly dependent on non-physical features ($>90\% $ correlation). In particular, the number of fibers in the IFU and the number of zeros in the maps show the strongest correlation with the group divisions in the PCA features.

\begin{deluxetable*}{lDDDDDDDD}
\tablenum{2}
\tablecaption{\label{tab:phys_depen}Cluster dependence with integrated parameters.}
\tablewidth{0pt}
\tablehead{\colhead{Cluster} & \multicolumn2c{$\log(\sigma)$} & \multicolumn2c{LW Age$_{R_{\rm e}}$} & \multicolumn2c{$\log(R_{\rm e}{\rm [kpc]})$} & \multicolumn2c{$\lambda_{R_{\rm e}}$} & \multicolumn2c{log(SFR$_{\rm H_{\alpha}}$)} & \multicolumn2c{LW ZH$_{R_{\rm e}}$} & \multicolumn2c{$\nabla$LW ZH$_{R_{\rm e}}$} & \multicolumn2c{$\nabla$LW Age$_{R_{\rm e}}$}}
\decimalcolnumbers
\startdata
1 & 13.4$\pm$3.7 \% & 35.2$\pm$1.6 \% & 14.6$\pm$1.4 \% & 17.9$\pm$1.9 \% & 6.8$\pm$0.6 \% & 4.5$\pm$0.8 \% & 6.4$\pm$0.3 \% & 1.1$\pm$0.3 \% \\
2 & 3.7$\pm$2.2 \% & 61.5$\pm$3.1 \% & 8.9$\pm$0.4 \% & 4.5$\pm$1.1 \% & 10.9$\pm$1.2 \% & 7.4$\pm$1.4 \% & 1.9$\pm$0.8 \% & 0.9$\pm$0.5 \% \\
3 & 43.6$\pm$1.6 \% & 14.2$\pm$1.6 \% & 11.2$\pm$1.2 \% & 16.7$\pm$1.5 \% & 1.3$\pm$0.5 \% & 4.9$\pm$0.7 \% & 3.9$\pm$0.4 \% & 4.0$\pm$0.3 \% \\
\enddata
\tablecomments{Dependence with integrated parameters of the three groups found with the $K$-means clustering in the SimCLR representation space.}
\end{deluxetable*}

The maximum number of clusters that fulfills our agreement and non-physical accuracy requirements is three. This number also corresponds with the natural divisions of the UMAP maps of Fig.~\ref{fig:2d_comparison} which clearly show a large cluster in the bottom left corner and a strong age trend in the right blob.

\begin{figure*}
   \centering
   \includegraphics[width=\hsize]{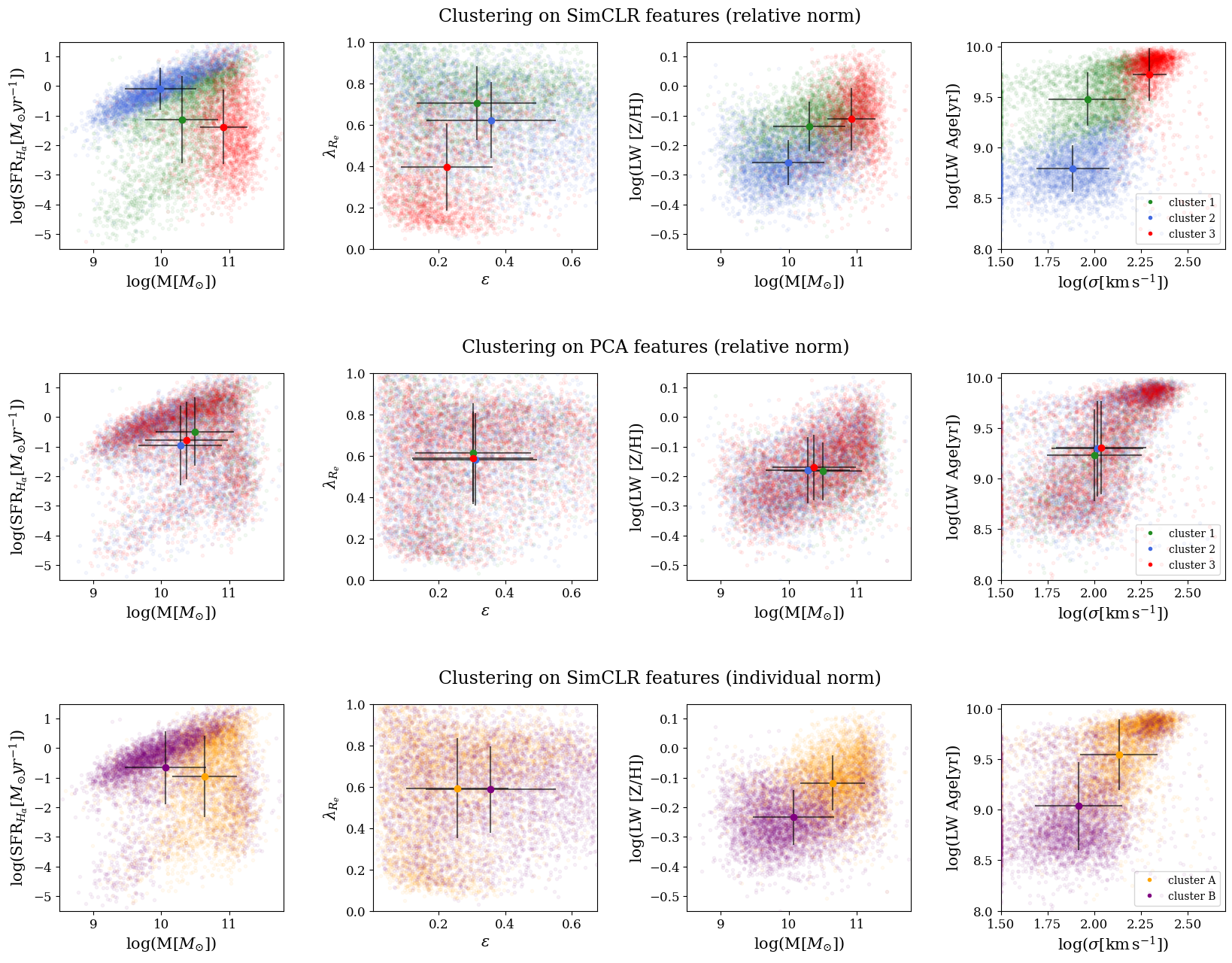}
      \caption{From left to right: stellar mass vs. SFR plane, ellipticity vs. specific angular momentum, stellar mass vs. LW metallicity, central velocity dispersion vs. LW age.  Top row panels show integrated parameters color-coded according to the 3 clusters found in the SimCLR representation space when the relative norm (see Sec.~\ref{sec:data}) is applied to the raw data. In the second row panels, the cluster division was performed on the PCA-1024 feature space. The bottom row shows the division into two clusters of the SimCLR representation space when the individial norm is applied to the input data. Mean values of each cluster are displayed in solid colors, errorbars account for the standard deviation of the cluster. Groups found with SimCLR representations show less overlap in all four physical planes than groups found by clustering PCA features.}
         \label{fig:3clus}
   \end{figure*}

In the following we only consider these three clusters and analyze their physical properties. Table~\ref{tab:phys_depen} shows the correlation of the clusters with several integrated physical parameters. The strongest correlations are with velocity dispersion, age, physical size and $\lambda_{R_{\rm e}}$. For a better visualization, we show in Fig.~\ref{fig:3clus} four well studied planes for galaxies: $\log M_*-\log {\rm SFR}$, $\epsilon-\lambda_R$, $\log M_*-[Z/H]$ and $\log\sigma-\log {\rm age}$.

We clearly observe that the three different clusters identified in a purely unsupervised way populate different regions in the different planes. While star-forming discs (with $\langle \log{\rm age [yr]} \rangle=9.9$ and $\langle \epsilon \rangle =0.36$) with low metallicity ($\langle Z/H \rangle=-0.25$) group in cluster 2, quenched fast-rotating galaxies with low to intermediate masses ($\log M_*/M_\odot\lesssim10.9$) can be found in cluster 1. Lastly, cluster 3 includes the massive ($\log M_*/M_\odot\gtrsim10.4$) slow-rotating galaxies. We emphasize that the star formation rate is not included in the maps used for the representation. Nevertheless, most of  main sequence galaxies are associated to one unique cluster which naturally suggests that star forming galaxies share similar kinematic and stellar population properties over several decades of stellar mass. We will discuss this further in Section~\ref{sec:disc}. For comparison, the bottom row of Fig.~\ref{fig:3clus} shows the same four physical planes obtained through clustering on the PCA space. Interestingly, we see that there is very little dependence on physical properties which again highlights the effectiveness in extracting physical information of the SimCLR representations when compared to those of the PCA.

\begin{figure}
   \centering
   \includegraphics[width=0.9\hsize]{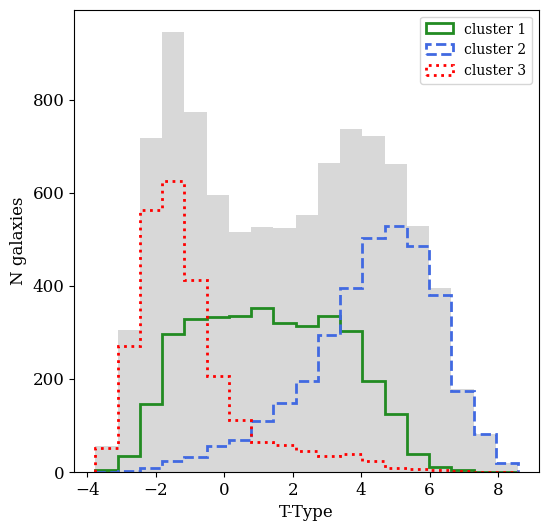}
      \caption{T-Type classification of all MaNGA galaxies (grey) and how they distribute across the unsupervised clusters (red, green and blue). The colors of the clusters have been chosen as the first row panels in Figure~\ref{fig:3clus}. Early type galaxies (T-Type$<0$) are predominantly in the cluster that includes the massive slow-rotating galaxies. Galaxies belonging the star forming main sequence cluster (blue) mostly distribute around the largest values of T-Type, while intermediate mass quenched galaxies show intermediate T-Type values.}
    \label{fig:3clus-morph}
   \end{figure}

Additionally, we analyze the morphological features of the galaxies in each cluster using the MaNGA Deep Learning DR-17 morphological catalogue (MDLM-VAC; private communication, Domínguez Sánchez et al. in prep). This catalogue is an extension of the  Deep learning DR-15 catalogue presented in \cite{Fischer-2019} and it was obtained with deep learning models trained on SDSS-DR7 images (see \cite{SDSS-Helena} for more details on the methodology and models performance). Figure~\ref{fig:3clus-morph} shows how the galaxies distribute in each cluster according to the morphological parameter T-Type.

We find that galaxies that have been classified as early types (T-Type\,$<0$) belong predominantly to the cluster containing high-mass slow-rotating galaxies, while a smaller fraction belongs to the intermediate-mass quenched galaxy cluster. Galaxies with T-Types\,$>0$ distribute in the clusters that include less massive galaxies, with the latest T-Types belonging to the star-forming cluster.


\section{Internal structure}
\label{sec:internal}

In the previous analysis the relative values of the observed maps were kept, since the normalization applied works as a uniform change of units across all the galaxy maps. We now analyze how the representation space organizes when only the spatial features of the input maps are considered. For this purpose, a linear normalization is applied for each map individually, such that the values of each map span the $[0, 1]$ range (individual norm). With this normalization, information about the absolute values of the different quantities is removed and the model is forced to focus exclusively on the internal structure of the observed galaxies. The training process is repeated on the newly normalized input data using the same augmentations and model settings. We present the analysis of the representation space performed as in Section~\ref{sec:res1}.

\subsection{Two-dimensional visualization of representations}

\begin{figure*}
   \centering
   \includegraphics[width=\hsize]{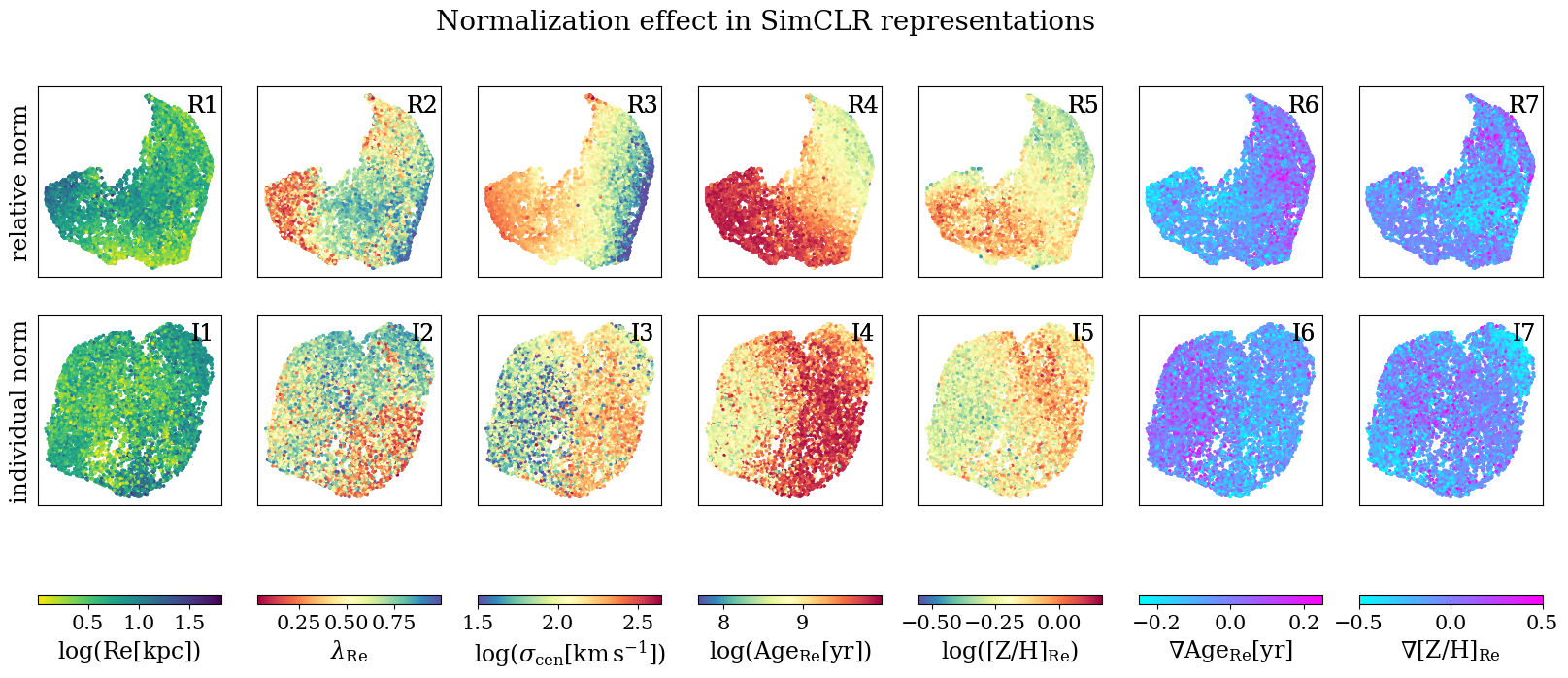}
      \caption{As Fig.~\ref{fig:2d_comparison}, 2D UMAP projection of SimCLR representations using relative norm (top row) and individual norm (bottom row). Color coding corresponds to physical properties derived from the maps, from left to right: effective radius, specific angular momentum, central velocity dispersion, luminosity weighted age and metallicity and their gradients. Although the individual norm removes the relative information of these parameters from the maps, the spatial features still correlate with them.}
         \label{fig:norms_integrated}
   \end{figure*}

As in Section~\ref{sec:2dvisual}, the SimCLR representations obtained after training the model are projected to a 2D space with UMAP \citep{umap} to visualize how the representation space correlates with known physical parameters. For comparison, the projections of both sets of normalized maps are presented together in Figure~\ref{fig:norms_integrated}.

While the relative norm representations show smoother transitions for central velocity dispersion, projected angular momentum, age and metallicity (panels R2, R3, R4 and R5), the correlation between these parameters and the individual norm representation space is not negligible. Furthermore, panels I3, I4 and I5 show that young and metal poor galaxies with low velocity dispersion tend to populate the left side of the maps, while old metal rich galaxies with higher velocity dispersion primarily locate on the right-hand side.

\subsection{Unsupervised Clustering of the representation space}

As the representation space distribution changes when applying different norms to the input data, we analyse the dominant drivers of the new space by performing an unsupervised clustering.

In this representation space, we find that only a division into two clusters fulfills the agreement requirement defined in Section~\ref{sec:clustering} with an agreement score of $90.89\pm0.03$, while more partitions yield scores $<78\%$. The non-physical parameter fits have comparable accuracy to the ones obtained with the relative norm. Specifically for the two cluster division, this value is $67.6 \pm 1.3$, which is only slightly larger than the $65\%$ threshold imposed in the previous section. To avoid confusion with the clusters of the previous section, we refer to the new groups as clusters $A$ and $B$.

The distribution of the two clusters the physical planes $\log M_*-\log {\rm SFR}$, $\epsilon-\lambda_R$, $\log M_*-[Z/H]$ and $\log\sigma-\log {\rm age}$ is shown on the bottom row of Figure~\ref{fig:3clus}. While cluster~$A$ is constrained in the planes towards higher mass, older and more metal rich regions (with average values $\langle \log{M_*/M_\odot} \rangle=10.6$, $\langle \log{age[yr]}\rangle=9.5$ and $\langle Z/H\rangle=-0.11$), cluster~$B$ spreads across all the planes with lower concentration in the overlapping regions (with average values $\langle \log{M_*/M_\odot}\rangle=10 $, $\langle \log{age[yr]}\rangle=9$ and $\langle Z/H\rangle=-0.23$). The low-mass and star forming galaxies are contained primarily in the last cluster. Although there is overlap between the two groups, a physical difference is captured and is clearly visible in age, mass and metallicity. This is interesting because the information about the absolute values has been removed, however, we still see that the main clusters found correlate with integrated properties such as stellar mass and velocity dispersion.

\begin{figure}
   \centering
   \includegraphics[width=\hsize]{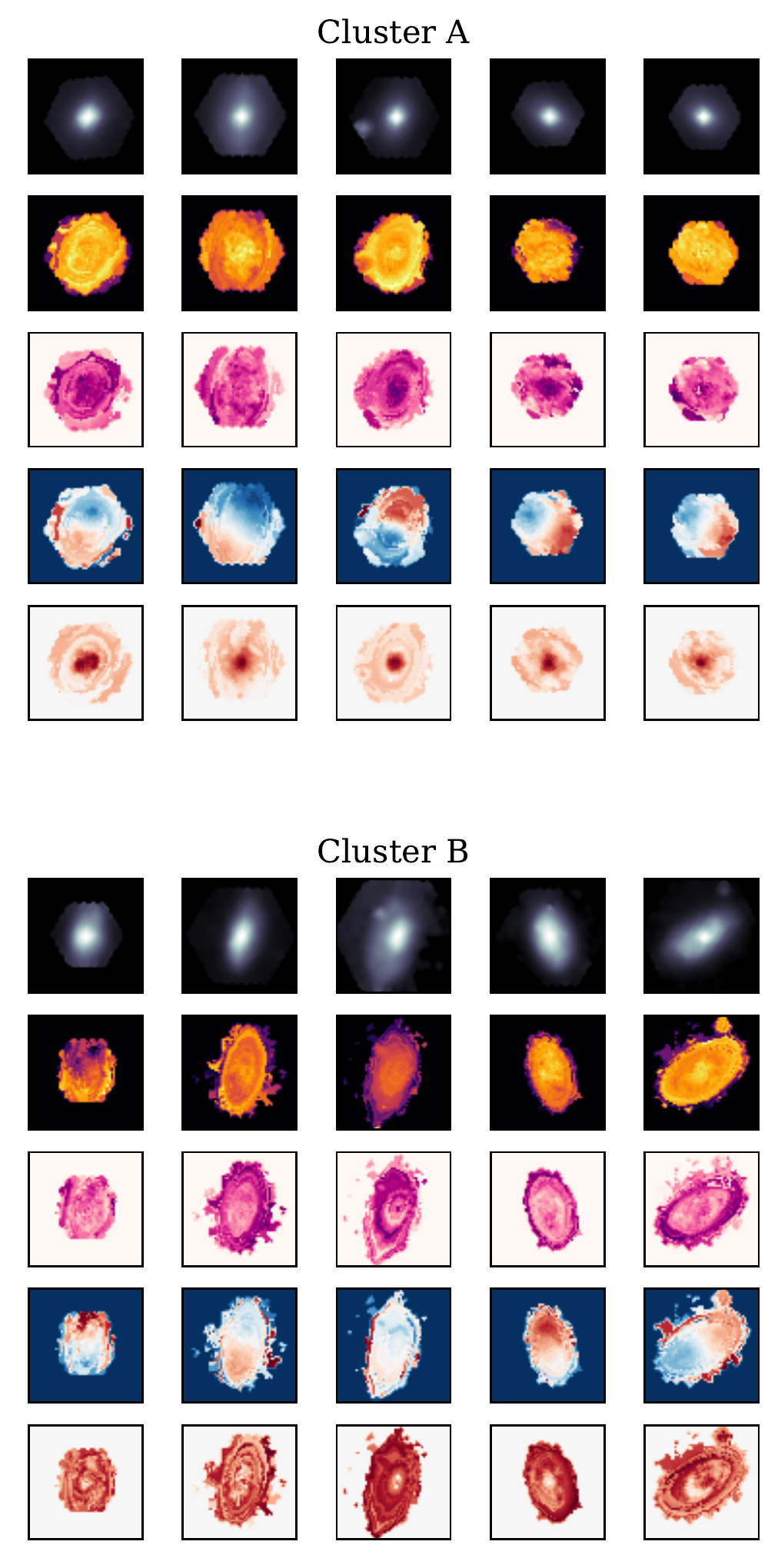}
      \caption{Example galaxies from clusters $A$ and $B$ (top and bottom panels, respectively). The V-band reconstructed image, age, metallicity, radial velocity and dispersion maps (rows) were normalized with the individual norm for each galaxy (columns).}
         \label{fig:best_examples_AB}
   \end{figure}

To further visualize the typical spatial features found in each group, we select the five galaxies whose representations are closer to the centroid of each cluster. The example galaxies of each cluster are shown in Figure~\ref{fig:best_examples_AB}. While the example galaxies from cluster $A$ show an early type morphology and negative gradients in velocity dispersion and metallicity maps, galaxies in cluster $B$ tend to have inverted gradients to those in cluster $A$ and their V-band reconstructed images show structure and disk features compatible with later type morphology.

As this division roughly separates early type galaxies from late types, the morphological features encoded in the V-band reconstructed images could be driving the clustering. Therefore, in Appendix~\ref{ap:no-morph} we evaluate the influence of the kinematic and composition maps in the division. When the V-band reconstructed images are excluded from the input maps, we find that the division can be recovered with good agreement ($88.621\pm0.007\%$ overlap with the previous clusters) indicating that the information of the four remaining maps is enough to find the two groups.
\section{Summary and Discussion}
\label{sec:disc}

\subsection{Unsupervised exploration of multi-dimensional datasets}

A key property of contrastive learning-based dimensionality reduction, in contrast with other techniques, is that representations are encouraged to become invariant to a set of transformations. We have shown that this feature has a lot of potential when the dataset is non-homogeneous and affected by known biases, since these can be tackled with convenient transformations in the contrastive framework. 

We confirm that the representations obtained in this work efficiently reduce their impact and focus on physical properties, even though MaNGA maps include important instrumental effects. This allows one to explore the physical correlations existing in the high dimensional space. 

Therefore, the presented framework might constitute an important tool to explore and discover patterns in future datasets. We have illustrated in this work how the representations organize based on five kinematics and stellar population maps. However, the method has no limit on the number of channels in the input data, thus it is possible to extend the analysis to a larger number of maps including for example gas properties or even complete IFU data cubes without post-processing.

Additionally, because transformations included in the training procedure can be tuned, one could eventually combine datasets of different origins and selection effects. A natural extension of this work is thus the combination of IFU data from different surveys (i.e. MUSE, \citealt{muse}; CALIFA, \citealt{califa}; MaNGA, \citealt{MaNGA-Bundy-2015}; SAMI, \citealt{sami}) represented in a unique low-dimension space. Combining observations and predictions from recent hydrodynamic cosmological simulations is also a promising avenue to identify possible discrepancies between the physical properties of observed and simulated galaxies. 

In addition to visualization, the representations learnt can be used in a variety of downstream tasks. \cite{chen2020big} and  \cite{selfsup-galaxy} found that this framework can obtain the same levels of accuracy for classification than a supervised approach but reducing the number of training labels. We presented in this work one application that makes use of the representation space as a tool to explore the dataset. 

\subsection{Clustering}

Since contrastive learning locates objects with similar physical properties in nearby regions of the representation space, it can be used to identify groups of objects with shared physical properties in a purely data driven manner. In Sections~\ref{sec:reps} and~\ref{sec:internal} we illustrated this procedure with two different normalization of the MaNGA maps: with and without information about the absolute values of the physical properties. This allows us to study the interplay between integrated properties of galaxies and their resolved internal structure.

\subsubsection{Absolute physical properties}

When applying the relative norm to the input data, the three groups identified naturally divide galaxies into three previously well-defined categories using integrated properties: main sequence rotating discs, low mass quiescent rotating galaxies and massive quiescent slow-rotating galaxies. This highlights the potential of our proposed framework for future discoveries. 

Although no direct star formation rate indicators were included in the input, the clustering algorithm naturally puts most of the main-sequence galaxies in the same group, i.e. the main sequence is \emph{rediscovered} in a data-driven way. It suggests that galaxies in the main sequence share kinematic and stellar populations properties and therefore evolve with similar physical processes despite covering several decades in stellar mass.

The well-known slow-rotating galaxies are primarily included in this group, which have been put forward by several previous IFU surveys (\citealt{lambda}, \citealt{Emsellem2011}, \citealt{Graham2018}, \citealt{Jesus2019}). The so-called slow rotators are thought to have an assembly history more dominated by major mergers \citep{cappellari-2016} which therefore increases their stellar mass and velocity dispersion. Our framework naturally isolates these objects.

There is a third interesting group highlighted in Fig.~\ref{fig:3clus} which is composed of low-mass quiescent galaxies. Interestingly, these galaxies are rotating fast as opposed to the slow rotators of the previous cluster. It suggests that the quenching mechanisms for this population is different. This could be predominantly a population of satellite galaxies quenched through strangulation processes \citep{Peng-Maiolino-2014} without significant impact on the kinematics of stars. 

While the three clusters found are well-known groups from previous morpho-kinematic works, our methodology found this classification with additional stellar population information: the luminosity weighted age and metallicity maps. In Appendix~\ref{ap:morpho-kin} we analyze if the clusters can be recovered by the framework when only the V-band image and kinematic maps are considered. While the clusters are highly dependent of these input maps, without the age and metallicity maps the classification cannot be fully recovered ($\sim25\%$ of the galaxies are misplaced in the new division). Furthermore, not only an expected smaller dependence with age and metallicity is found in the average cluster properties, but a greater overlap of the clusters in T-Type classification is noticeable. We note, however, that the massive early-type galaxies tend to be more consistently classified in the same group, again highlighting that the distinctive kinematic features of these galaxies isolate them from the rest. Interestingly, removing the V-band reconstructed image form the input data and including the four remaining maps has no significant effect on the T-Type classification per cluster (see Appendix~\ref{ap:no-morph}). 

We conclude that the divisions that naturally arise from this framework confirm extensively studied groups of galaxies. The clusters found reflect the dominant differences between three regions of the representation space. The clusters share main properties with well-known galaxy categories, confirming that these groups of galaxies present intrinsic differences. This again highlights the potential of our proposed framework to discover the physical mechanisms that shape galaxies.

The results might, however, still be affected by selection effects. The MaNGA sample is not a volume-limited sample and this is not taken into account in the representations. Although a set of weights to obtain a statistically representative sample of a single volume are provided in \cite{Wake_2017}, limiting the sample would mean a significant reduction of the number of galaxies available to train the Deep Learning model. This may conflict with the method as Deep Learning models require a large training set to avoid an overfit of the network's parameters. Although the different groups do exist, their relative fractions or relevance might not be representative of the local Universe. Also small groups of galaxies might not be detected.

\subsubsection{Internal structure}

\cmmnt{}

We have repeated the cluster analysis with a different normalization that removes information about the absolute values of the physical properties. This allows us to quantify the relation between integrated and resolved internal structure of nearby galaxies.   Although the previous clusters cannot be completely recovered with this new normalization, it is possible to find two clusters with good agreement that are separated primarily by the spatial features present in the maps. These clusters correlate with the mean and integrated physical values derived from the maps, suggesting that the internal structure of galaxies serve as a footprint that is tightly related to the observed integrated properties. It suggests that these two galaxy types are different not only because they differ in the absolute values, but also because their internal structure is different, pointing towards distinct assembly histories, as highlighted in previous works.

The first cluster found when only spatial features are considered contains intermediate to high-mass quenched galaxies. These galaxies show in general  early type morphologies and present negative slopes in the velocity dispersion and metallicity maps (Figure~\ref{fig:best_examples_AB}). The second cluster primarily includes the star forming main sequence of galaxies. Figure~\ref{fig:best_examples_AB} shows that galaxies in this cluster tend to present flat metallicity and velocity dispersion distributions and more disky like stellar structure.

\begin{figure}
   \centering
   \includegraphics[width=\hsize]{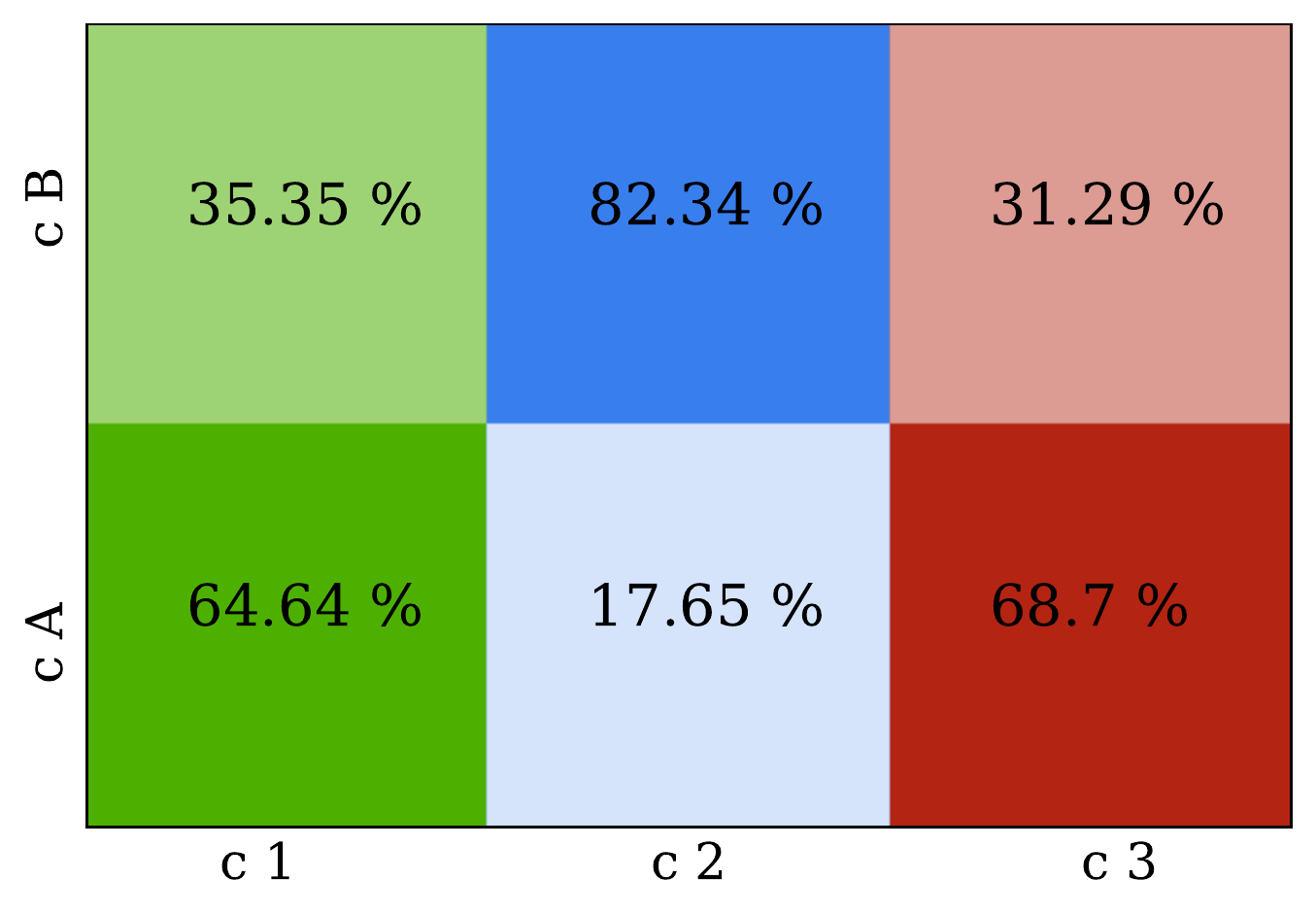}
      \caption{Overlap of clusters found in using relative (clusters $1$,$2$ and $3$ - vertical) and individual norms (clusters $A$ and $B$ - horizontal). Percentages are calculated over the relative norm clusters.}
         \label{fig:confusion_overlap}
   \end{figure}

The comparison between the clusters obtained with the different normalizations reveals interesting trends. Figure~\ref{fig:confusion_overlap} shows that $\sim80\%$ and $\sim70\%$ of the galaxies that were in different clusters when including the information about the absolute values are also separated by internal structure. It confirms that the distinction between high velocity and low velocity dispersion systems reveals different assembly histories and it is not only a reflection of a stellar mass bimodality. Interestingly, the third cluster found in the  relative normalization run is distributed among the two main ones with $\sim35\%$ being closer to the star-forming rotating disks and $\sim65\%$ being more similar to the massive slow rotating systems. It suggests that this intermediate mass quenched population is formed through a mixture of smooth and more violent assembly processes but there is no a particular signature in the internal kinematics of the galaxies that clearly distinguishes them from the two other groups. A fraction of the objects also change cluster when the normalization is modified. We will study these systems in more detail in future work.

\section{Conclusions}

We applied a method based on the SimCLR deep contrastive learning approach for obtaining meaningful representations of galaxy resolved maps from the MaNGA survey. The method produces a low-dimension representation of the data input while preserving meaningful physical information and reducing the impact of the instrumental effects present in the data. 

We showed that the proposed representation learning scheme outperforms, at least for this particular dataset, more standard dimensionality reduction algorithms such as PCA which is more sensitive to non-physical correlations. 

We have shown that the representation space can be efficiently used for visualization, clustering and exploring the properties of sub-categories of objects. We argue that contrastive learning techniques represent a promising avenue to explore and discover patterns in future high-dimensional astronomical datasets.

\begin{acknowledgements}

We thank Sara Ellison, Mallory Thorp and David Patton for comments on a previous version of this manuscript. We acknowledge financial support from the European Union's Horizon 2020 research and innovation programme under Marie Sk\l odowska-Curie grant agreement No 721463 to the SUNDIAL ITN network, from the State Research Agency (AEI-MCINN) of the Spanish Ministry of Science and Innovation under the grant "The structure and evolution of galaxies and their central regions" with reference PID2019-105602GB-I00/10.13039/501100011033, and from IAC project P/300724, financed by the Ministry of Science and Innovation, through the State Budget and by the Canary Islands Department of Economy, Knowledge and Employment, through the Regional Budget of the Autonomous Community. J.~F-B  acknowledges support through the RAVET project by the grant PID2019-107427GB-C32 from the Spanish Ministry of Science, Innovation and Universities (MCIU), and through the IAC project TRACES which is partially supported through the state budget and the regional budget of the Consejer\'ia de Econom\'ia, Industria, Comercio y Conocimiento of the Canary Islands Autonomous Community.

This project makes use of the MaNGA-Pipe3D dataproducts. We thank the IA-UNAM MaNGA team for creating this catalogue, and the Conacyt Project CB-285080 for supporting them.

Funding for the Sloan Digital Sky 
Survey IV has been provided by the 
Alfred P. Sloan Foundation, the U.S. 
Department of Energy Office of 
Science, and the Participating 
Institutions. SDSS-IV acknowledges support and 
resources from the Center for High 
Performance Computing  at the 
University of Utah. The SDSS 
website is www.sdss.org.
SDSS-IV is managed by the 
Astrophysical Research Consortium 
for the Participating Institutions 
of the SDSS Collaboration including 
the Brazilian Participation Group, 
the Carnegie Institution for Science, 
Carnegie Mellon University, Center for 
Astrophysics | Harvard \& 
Smithsonian, the Chilean Participation 
Group, the French Participation Group, 
Instituto de Astrof\'isica de 
Canarias, The Johns Hopkins 
University, Kavli Institute for the 
Physics and Mathematics of the 
Universe (IPMU) / University of 
Tokyo, the Korean Participation Group, 
Lawrence Berkeley National Laboratory, 
Leibniz Institut f\"ur Astrophysik 
Potsdam (AIP),  Max-Planck-Institut 
f\"ur Astronomie (MPIA Heidelberg), 
Max-Planck-Institut f\"ur 
Astrophysik (MPA Garching), 
Max-Planck-Institut f\"ur 
Extraterrestrische Physik (MPE), 
National Astronomical Observatories of 
China, New Mexico State University, 
New York University, University of 
Notre Dame, Observat\'ario 
Nacional / MCTI, The Ohio State 
University, Pennsylvania State 
University, Shanghai 
Astronomical Observatory, United 
Kingdom Participation Group, 
Universidad Nacional Aut\'onoma 
de M\'exico, University of Arizona, 
University of Colorado Boulder, 
University of Oxford, University of 
Portsmouth, University of Utah, 
University of Virginia, University 
of Washington, University of 
Wisconsin, Vanderbilt University, 
and Yale University.

\end{acknowledgements}

%
%
\bibliography{biblio}{}
\bibliographystyle{aasjournal}

\appendix

To analyze the dependence of the representation space with the input maps we repeat the clustering analysis with different combinations of maps. In particular, we consider two variations to the proposed methodology that tackle specific questions. On the one hand, we examine the cluster division when morphology information is excluded. For this purpose, we perform $K$-means clustering on a representation space obtained by excluding the V-band reconstructed image. On the other hand, we analyze if the three clusters found in Section~\ref{sec:clustering} can be recovered solely from the morpho-kinematic input data.

\section{Age, metallicity and kinematics}\label{ap:no-morph}

\begin{figure}[h]
\centering
\begin{minipage}[b]{.45\linewidth}
    \includegraphics[width=\linewidth]{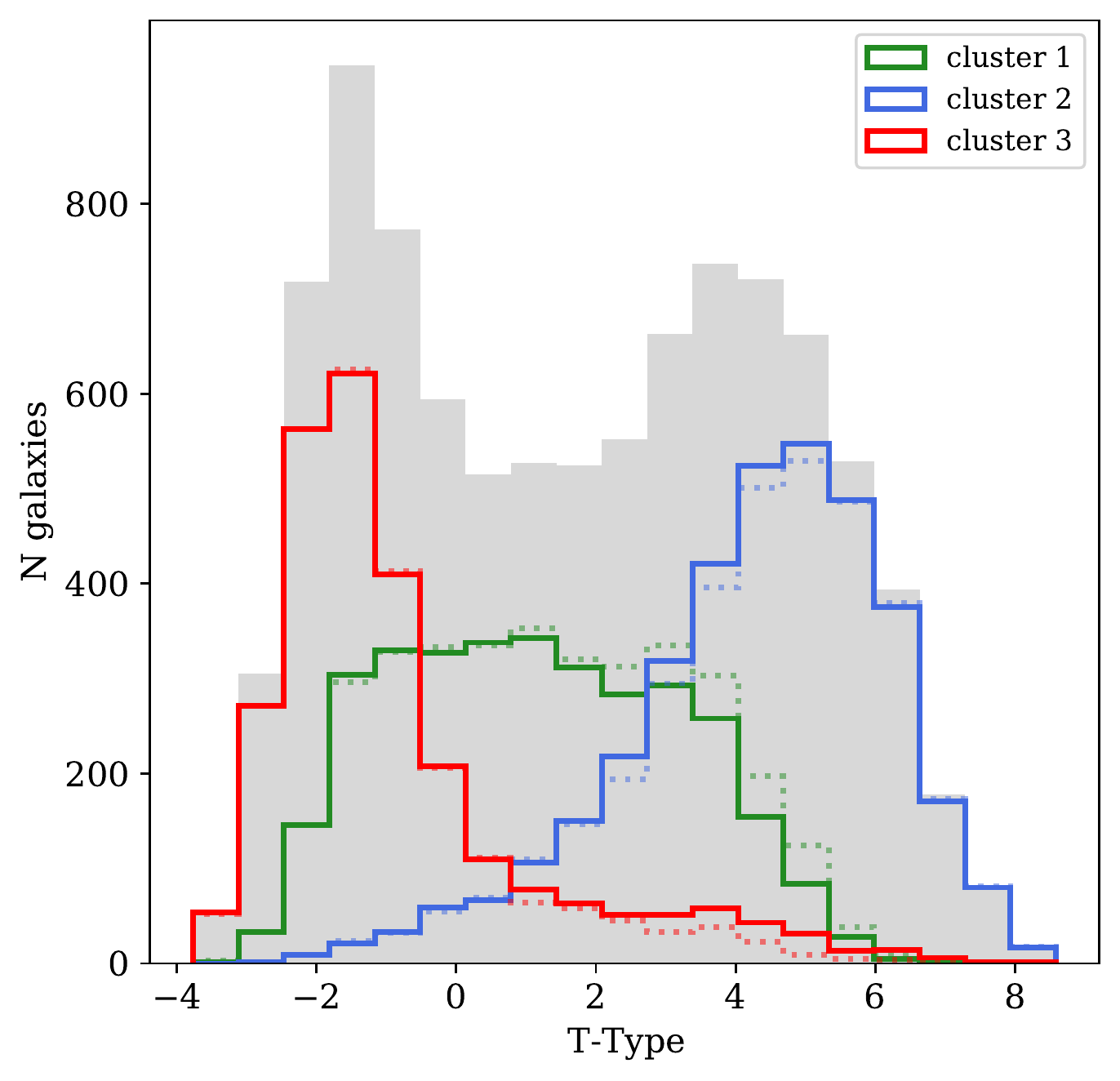}
    
\end{minipage}
\quad
\begin{minipage}[b]{.45\linewidth}
    \includegraphics[width=\linewidth]{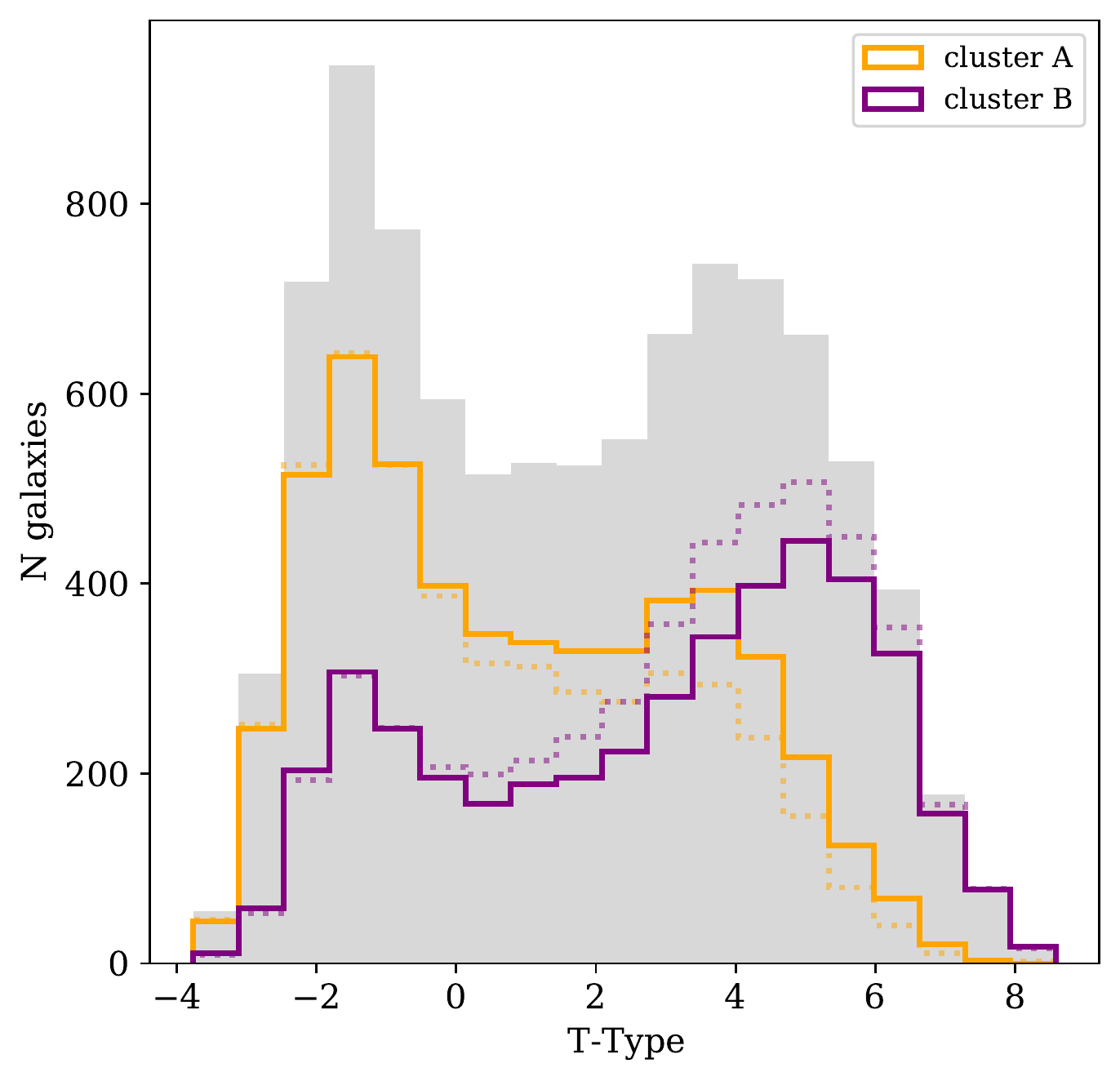}
\end{minipage}
\caption{As Figure~\ref{fig:3clus-morph}, the grey filled histograms show the T-type classification of all MaNGA galaxies and colored line histograms correspond to unsupervised clusters. Spot line histograms correspond to clusters obtained with the five maps as input: V-band reconstructed image, age, metallicity, radial velocity and velocity dispersion maps. Solid lines show the cluster divisions when the the V-band reconstructed image is excluded. Left panel: colors correspond to the 3 clusters found when the \emph{relative} norm is applied to the input data. Right panel: colors correspond to the 2 clusters found when the \emph{individual} norm is applied. In both panels the clusters' trends are maintained after removing the V-band image from the input. The discrepancy between clusters is clearer toward later-type galaxies.}
\label{fig:5to4maps_overlap}
\end{figure}

We train the SimCLR model on the age, metallicity, radial velocity and dispersion maps. We later use this trained model to extract the representations of this set of maps and perform $K$-means on them. When the input data is normalized with the relative norm the clusters overlap a $92.04\pm0.03\%$ (error estimated as the standard deviation of ten different $K$-means initialisation). This value is slightly smaller than the agreement score for the clustering on the representation space obtained  when the V-band reconstructed image is included, and therefore this channel does contribute to the $3$ cluster division. We find that the clusters show more discrepancy in the later-type galaxies (T-Type\,$>\,0$, as seen the left panel of Figure~\ref{fig:5to4maps_overlap}). However, we conclude that the overlap is significant since it is indeed greater than the  agreement threshold ($85\%$) defined in Section~\ref{sec:clustering}.

When this analysis is repeated on the maps normalized with the individual norm, an overlap of $88.621\pm0.007\%$ is recovered. After removing the V-band image, the clusters tend to equalize their T-Type distributions in later type galaxies (Fig.~\ref{fig:5to4maps_overlap}). However, the agreement of the clusters before and after excluding the V-band image is significant and the overall trends of both clusters are still clear. This indicates that the division does not only depend of the galaxy morphology, but the remaining maps have sufficient information to separate the two groups.

\section{Morpho-kinematics}\label{ap:morpho-kin}

While the three clusters found in Section~\ref{sec:clustering} are well known groups from previous morpho-kinematic works, our methodology found these classification with additional stellar population information: the luminosity weighted age and metallicity maps. Therefore, we analyze if the clusters can be recovered by the framework when only the morpho-kinematic maps are considered.

The clusters that arise when the age and metallicity maps are excluded overlap in a $76.95\pm0.02\%$ with those clusters found in Section~\ref{sec:clustering}. While the morpho-kinematic maps greatly influence the clusters found in Section~\ref{sec:clustering}, we consider that the initial groups cannot be fully recovered because $\sim25\%$ of the galaxies are misplaced when the age and metallicity maps are excluded.

\begin{figure}
   \centering
   \includegraphics[width=\hsize]{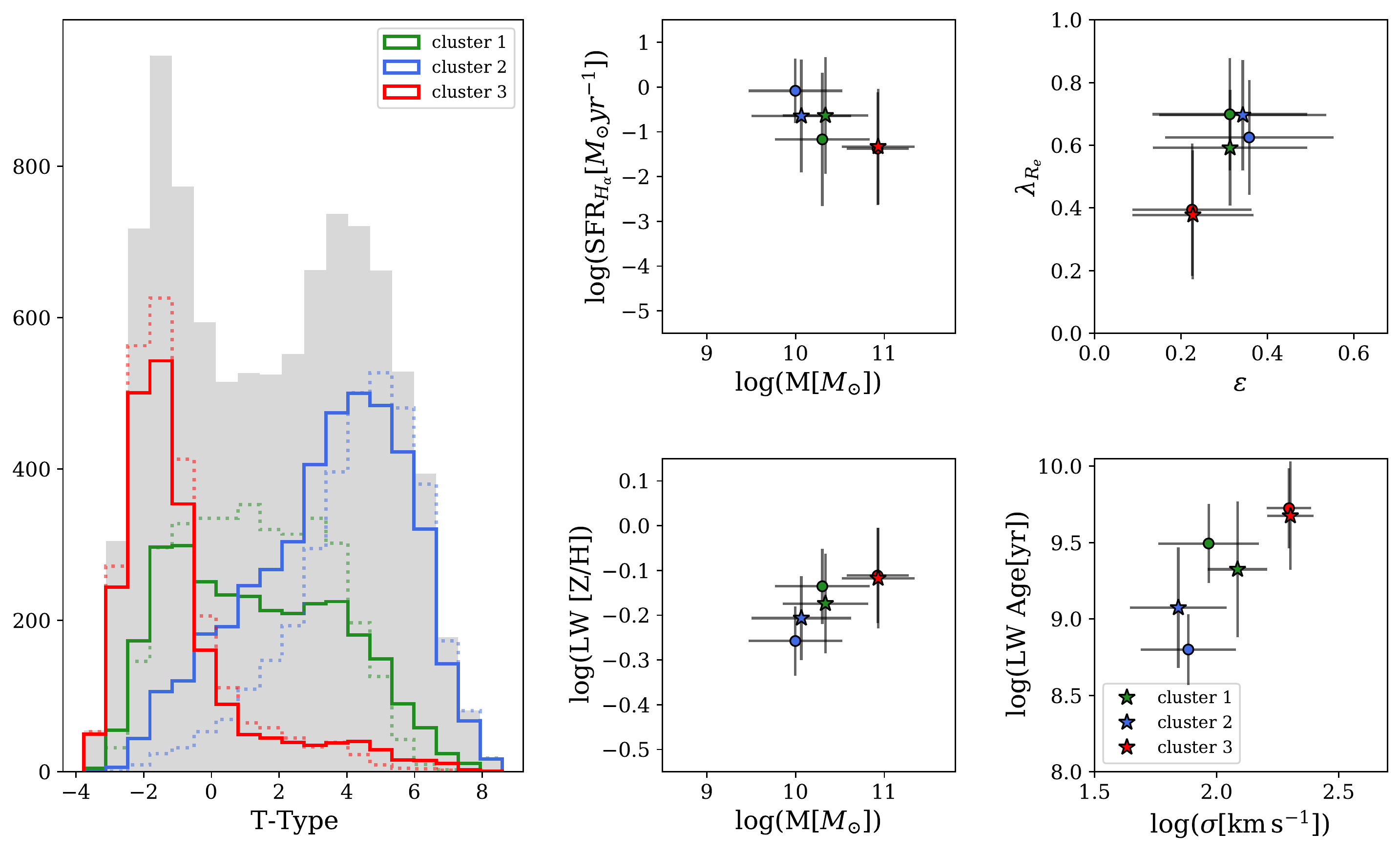}
      \caption{Left panel: T-Type classification. As Fig.~\ref{fig:5to4maps_overlap}, spotted line histograms correspond to clusters obtained when five maps are used as input. Solid line histograms . Small-sized panels on the right-hand side show the average integrated properties of the clusters. 5-map clusters are identified with circle markers, while star markers where used for the morpho-kinematic data.}
         \label{fig:morpho-kin_clusters}
   \end{figure}

We further inspect how the average properties of each cluster vary in Figure~\ref{fig:morpho-kin_clusters}. While cluster $3$ is the least affected of all three groups, clusters $1$ and $2$ show more overlap in T-Type classification and their mean group properties tend to equalize. However, there are parameters that show a larger separations in the physical planes (center and right panels of Fig.~\ref{fig:morpho-kin_clusters}), like $\sigma_{\rm cen}$ and $\lambda_{\rm Re}$. The fact that the difference between the clusters is enhanced only for these parameters is consistent with these parameters being derived from the maps used in this test while excluding the age and metallicity information. These results are consistent with the correlations analyzed in Table~\ref{tab:phys_depen}, which suggest that the cluster with the most massive galaxies is primarily separated due to their kinematics, while the low to intermediate mass galaxies are separated into old/metal-rich and young/metal-poor groups.

\end{document}